\documentclass[aps,prb,twocolumn,showpacs,preprintnumbers,
amsmath,amssymb]{revtex4}
\usepackage{graphicx}
\usepackage{dcolumn}
\usepackage{verbatim}
\def\mc{\multicolumn}

\begin{document}

\title{First-principles study of the lattice and electronic structures of 
TbMn$_2$O$_5$}

\author{Chenjie Wang}
\affiliation{Key Laboratory of Quantum Information, University of
Science and Technology of China, Hefei, 230026, People's Republic of
China}

\author{Guang-Can Guo}
\affiliation{Key Laboratory of Quantum Information, University of
Science and Technology of China, Hefei, 230026, People's Republic of
China}

\author{Lixin He
\footnote{corresponding author, Email address: helx@ustc.edu.cn}}
\affiliation{Key Laboratory of Quantum Information, University of Science and
Technology of China, Hefei, 230026, People's Republic of China}

\date{\today}

\begin{abstract}

The structural, electronic and
lattice dielectric properties of multiferroic
TbMn$_2$O$_5$ are investigated using density functional theory 
within the generalized gradient approximation (GGA).
We use collinear spin approximations and ignore the spin-orbit coupling.
The calculated structural parameters are in excellent
agreement with the experiments.
We confirm that the ground state structure
of TbMn$_2$O$_5$ is of space group $Pb2_1m$, allowing polarizations
along the $b$-axis.
The spontaneous electric polarization is calculated 
to be $1187$ $nC\cdot$cm$^{-2}$.
The calculated zone-center optical phonons frequencies 
and the oscillator strengths of IR phonons
agree very well with the experimental values.
We then derive an effective Hamiltonian to explain the magnetically-induced
ferroelectricity in this compound.
Our results strongly suggest that the ferroelectricity in TbMn$_2$O$_5$ is
driven by the magnetic ordering that breaks the the inversion symmetry,
without invoking the spin-orbit coupling.

\end{abstract}
\maketitle


\section{Introduction}

Recently, a large class of manganese
oxides (RMnO$_3$, \cite{kimura03,goto04} 
and RMn$_2$O$_5$, \cite{hur04,chapon04,blake05} 
with R=Y, Tb, Dy, etc.) has been discovered to
be multiferroic, with strong magnetoelectric (ME)
coupling. 
The ME coupling leads to various novel physical effects,
such as the ``colossal magnetodielectric'' (CMD) effects  
and magneto-spin-flop effects. \cite{kimura03,goto04,hur04b}
For example, in TbMn$_2$O$_5$,\cite{hur04,cheong07} 
the ME coupling is so strong, that the electric polarization 
can be reversed by applying a magnetic field. \cite{hur04}
The remarkable ME effects revealed 
in these materials have attracted great attention 
\cite{fiebig05,kagomiya03,kimura03,goto04,blake05,hur04, 
chapon04,aguilar06,cheong07,katsura05, sergienko06}   
because of the fascinating physics
and their potential applications in novel multifunctional 
ME devices.

\begin{figure}
\centering
\includegraphics[width=3.in]{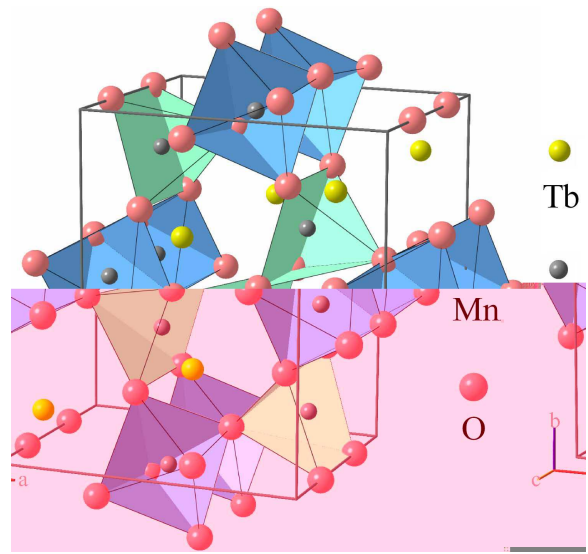}
\caption{(Color online) Structure of TbMn$_2$O$_5$ unit cell, 
showing Mn$^{4+}$O$_6$ Octahedra and  Mn$^{3+}$O$_5$ pyramids.
} 
\label{fig:structure}
\end{figure}

The crystal structure of TbMn$_2$O$_5$ is orthorhombic, 
with four chemical formula units per primitive cell 
(32 atoms in total), containing Mn$^{4+}$O$_6$ octahedra and 
Mn$^{3+}$O$_5$ pyramids, as shown in Fig.\ref{fig:structure}.
TbMn$_2$O$_5$ shows several magnetic phase transitions 
accompanied with the
appearance of electric polarizations and dielectric anomalies,
when cooling down from the room temperature. \cite{hur04, chapon04, blake05}
Starting from an incommensurate antimagnetic (ICM) ordering at $T_N$
= 43K with a propagation vector ${\bf k}$ ($\sim$0.50, 0, 0.30), 
the structure locks into commensurate antimagnetic (CM) 
state at $T_{CM}$ = 33K with
$\bf{k}$=(0.5, 0, 0.25), during which spontaneous polarization arises at
$T_{FE}$ = 38K. \cite{hur04, chapon04}
When the temperature lowers to $T_{ICM}$ = 24 K, 
magnetic order becomes ICM again, with a sudden decrease
of polarization and a jump of the $\bf{k}$ vector to (0.48, 0, 0.32).
The spontaneous polarization increase again,
when continuing to cool to about 10K. \cite{hur04}
During the magnetic phase transitions,
a peak at $T_{FE}$ and a step at $T_{ICM}$ of the dielectric constant was
observed,\cite{hur04, chapon04} indicating strong ME coupling in
this compound.
It was demonstrated the electric polarization can be
reversed by applying magnetic field. \cite{hur04}
 
Experimental data show that 
the structure of TbMn$_2$O$_5$
has space group \emph{Pbam}, \cite{alonso97} which
includes spatial inversion ($R^{-1}$) symmetry.
It is therefore puzzling that the material can develop spontaneous
electric polarizations.
It has been suspected \cite{kagomiya03, chapon04}
that the actual symmetry group of TbMn$_2$O$_5$
is \emph{Pb}2$_1$\emph{m}, allowing polarization along the $b$ axis.
Indeed, there are several experiments supporting this hypnosis.
\cite{chapon04,blake05,aguilar06} For example,
some Raman modes were found to be IR active in TbMn$_2$O$_5$, \cite{aguilar06}
and the anomalies of atomic displacement parameters (ADP)
have been observed.\cite{chapon04} 
Nevertheless, no {\it direct}
evidence of the lower symmetry has yet been found. \cite{chapon04, blake05}

Theoretically, the microscopic origin of the strong ME coupling and
the electric polarization is still under intensive debates.
\cite{chapon04, cheong07,katsura05, sergienko06}
The ME coupling could originate either from the symmetric
superexchange interactions, or from the antisymmetric exchange
interactions. \cite{fiebig05}
The antisymmetric exchange comes from the spin-orbit coupling, and the  
noncollinearity of the spin structure 
is an essential ingredient for this mechanism.
\cite{fiebig05,katsura05, sergienko06,hu07}
However it was shown in Ref. \onlinecite{chapon04} that the largest electric
polarization in TbMn$_2$O$_5$ 
is associated with the CM state that is
almost collinear. \cite{chapon06} 
In our recent work, \cite{wang07} 
we determined the ground-state structure of 
TbMn$_2$O$_5$ using the first-principles methods. 
The results show that the ground-state structure is indeed
of polar space group $Pb2_1m$ and the electric polarization equals 
$1187$ $nC\cdot$cm$^{-2}$.
In the calculations, we use collinear spin approximation and ignore the
spin-orbit interaction, suggesting that ME coupling in TbMn$_2$O$_5$
is due to the symmetric superexchange interactions.

The aim of the present work to examine rigorously
the ground state structural, electronic and
lattice dielectric properties 
of TbMn$_2$O$_5$ using first-principles calculations 
to provide solid
ground for further investigations.
The rest of paper is organized as follows.
After a brief discussion of the first-principles methods
and the approximations used in the calculations 
in Sec. \ref{sec:methodology},
we provide a detailed analysis of the ground-state
structural and electronic properties 
in Sec. \ref{sec:structure}, \ref{sec:e_structure}.
In Sec. \ref{sec:phonon}, 
we calculate all zone center optical 
phonon frequencies and the oscillator strengths of IR modes. The results are
in excellent agreement with the known experimental IR and Raman spectra.
In Sec. \ref{sec:polarization}, 
we calculate electric polarization in TbMn$_2$O$_5$.
We then derive an effective Hamiltonian to explain the 
microscopic mechanisms of
the ferroelectricity and the giant magnetoelectric coupling.
We conclude in Sec. \ref{sec:summary}.

\section{Methodology}
\label{sec:methodology}

Our calculations are based on the 
standard density-functional (DFT) theory with
spin-polarized generalized gradient approximation (GGA).
We adopt Perdew-Burke-Ernzerhof functional\cite{perdew96} implemented
in the Vienna \emph{ab initio} Simulations Package (VASP).
\cite{kresse93,kresse96} A plane-wave basis and projector
augmented-wave (PAW) pseudopotentials \cite{blochl94} are used, with
Mn 3\emph{p}3\emph{d}4\emph{s}, and  Tb 5\emph{p}5\emph{d}6\emph{s} 
electrons treated self-consistently.
A 500 eV plane-wave cutoff results in good convergence of the total
energies. We relax the structure until the changes of total energy
in the self-consistent calculations are less than 10$^{-7}$ eV, and
the remaining forces are less than 1 meV/\AA. 
Experimentally, TbMn$_2$O$_5$ is found to be incommensurate
anti-ferromagnetic (AFM) below 24 K, 
with the propagation vector ${\bf k} \approx (0.48, 0, 0.32)$. 
To accommodate the magnetic structure, one needs a
huge supercell, which is computationally prohibitive.
Instead, we use a 2$\times$1$\times$1 supercell,
equivalent to approximating the propagation vector ${\bf k}=(0.5, 0, 0)$.
The validity of this approximation has been
justified in our previous work. \cite{wang07}
For the supercell we used, a
$1\times2\times4$ Monkhorst-Pack k-points mesh converges very well the results.

It was demonstrated in Ref. \onlinecite{chapon04} that the largest electric
polarization is associated with a commensurate magnetic (CM) state that is
almost collinear. \cite{chapon06} 
Therefore, in the calculations, we use the
collinear spin approximation and ignore the spin-orbit coupling.
Our results agree very well with the known experiments,
indicating that these approximations capture the essential physics in
TbMn$_2$O$_5$.

\begin{table}
\begin{center}
\tabcolsep 5mm \caption{The calculated lattice constants compared with
experimental data. $FM$ and $L$ are the
structures of spin configurations $a$ and $g$ in
Fig. \ref{fig:spinconfigurations},
 respectively.}
\begin{tabular}{c c c c}
\hline \hline
lattice const. (\AA).  & $FM$ & $L$ & Exp.\cite{alonso97} \\
\hline a &7.3170  &7.3014  &7.3251\\
 b       &8.5269  &8.5393  &8.5168\\
 c       &5.6611  &5.6056  &5.6750\\
\hline
\end{tabular}
\label{tab:lattice_const}
\end{center}
\end{table}

\begin{figure}
\centering
\includegraphics[width=3in]{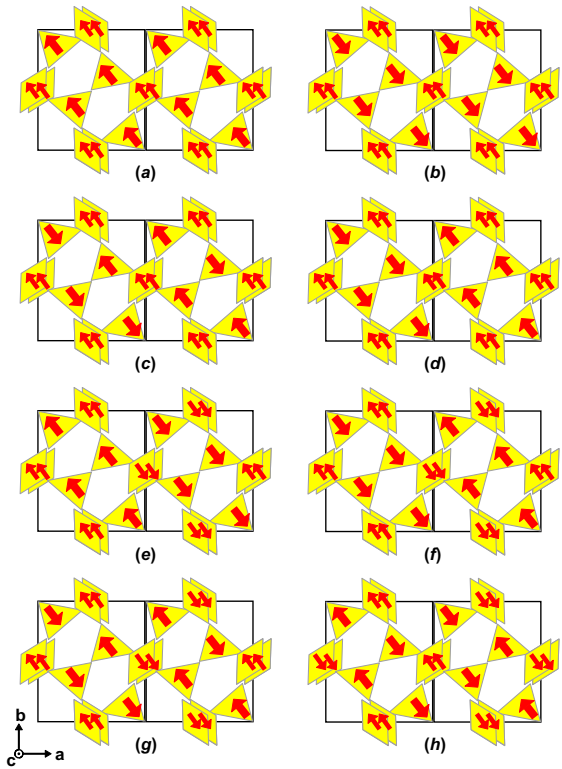}
\caption{(Color online) Spin configurations of
Mn$^{3+}$, Mn$^{4+}$ ions in the $ab$ plane.  
} 
\label{fig:spinconfigurations}
\end{figure}

\begin{figure*}
\centering
\includegraphics[width=4.5in]{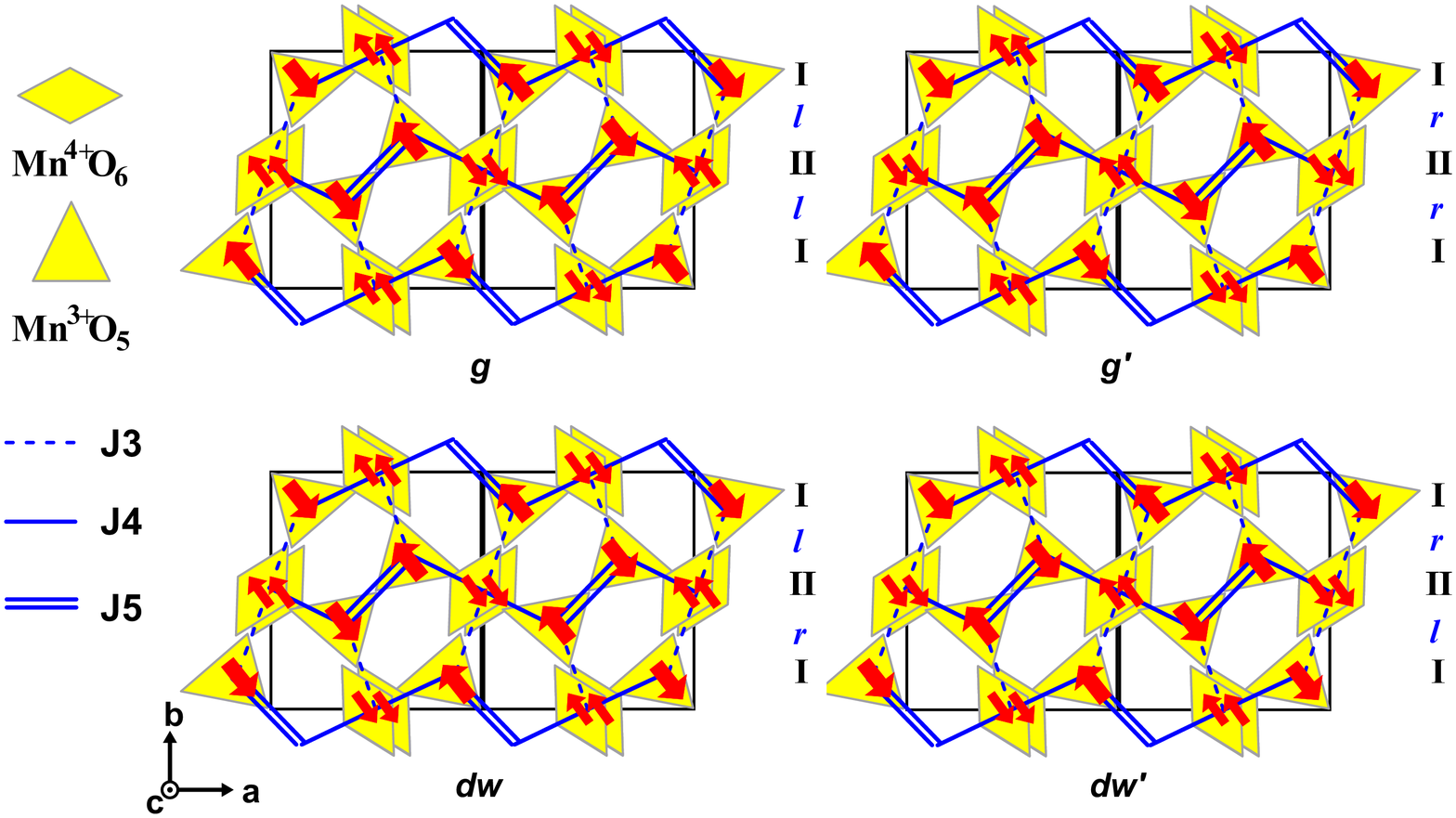}
\caption{(Color online) Spin configurations with
different combinations of chains I and II. 
$g$ and $g'$ are the ground state spin configurations for structure $L$ and
$R$ respectively, whereas 
$dw$ and $dw'$ are the domain walls between
structure $L$ and $R$. } 
\label{fig:chains}
\end{figure*}

\section{Ground-state Structure}
\label{sec:structure}

We start the structural relaxation from the experimental 
structural parameters. \cite{alonso97}
To determine the ground state structure, we tried various spin configurations
(SCs). 
Eight selected SCs, four ferromagnetic/ferrimagnetic (FM) and four
antiferromagnetic (AFM), are shown in Fig.\ref{fig:spinconfigurations}.
Here, we consider only the spins of Mn$^{3+}$ and Mn$^{4+}$ ions. 
The total energy of each SCs is calculated with full relaxations of the
(electronic and lattice) structures.
The stablest SC, i.e., the SC of lowest total energy, 
labeled as $g$ in Fig.\ref{fig:spinconfigurations} 
is AFM and is identical to the spin structure
proposed in Ref.\onlinecite{chapon04}. 
SC $g$ has an energetically degenerate SC,\cite{wang07} 
labeled as $g'$ in Fig. \ref{fig:chains}.
In both SCs $g$ and $g'$, Mn$^{4+}$ form an 
AFM square lattice in the $ab$ plane, whereas Mn$^{3+}$ couples to
Mn$^{4+}$ either antiferromagnetically via $J_4$ along $a$ axis or
with alternating sign via $J_3$ along the $b$ axis.
Mn$^{3+}$ ions in two connected pyramids also couple
antiferromagnetically through $J_5$. Here, we adopt the notations
$J_3$, $J_4$ and $J_5$ from Ref. \onlinecite{chapon04}, and define
the $J_3$ to be the Mn$^{4+}$- Mn$^{3+}$ superexchange interaction
through pyramidal base corners, and $J_4$ the superexchange
interaction through the pyramidal apex, as indicated in 
Fig. \ref{fig:chains}. We label the two different Mn$^{4+}$ chains
along the $a$ axis I, II, respectively, also following Ref.
\onlinecite{chapon04}.
The magnetic structure of $g'$ can be obtained from $g$ by shifting
chain II to the right (or to the left) by one unit cell along the
$a$ axis. \cite{chapon04} 
The exchange integrals $J_3$, $J_4$, $J_5$, were fitted
via a Heisenberg model assuming nearest neighbor (NN)
coupling.
We find that all the exchange energies are of AFM type, 
i.e., negative and
$|J_4|, |J_5| \gg |J_3|$. 
Therefore, the spins must couple via $J_4$ and $J_5$ anti-ferromagnetically 
in the stable magnetic structures.
More details about the spin structure is given in 
Appendix \ref{sec:appendix}.

The lattice structure relaxed from SC $g$ lead to structure $L$ in
Ref. \onlinecite{wang07}, whereas SC $g'$ gives the 
structure $R$.
The calculated lattice constants \cite{wang07} 
of the ground state structure $L$ (and
$R$) are listed in Table \ref{tab:lattice_const},
and are in very good agreement with the experiments. \cite{alonso97}
The errors of the lattice constants are about 1\%, 
typical errors for GGA. 
The lattice constants calculated for the full FM configuration (SC $a$),
are also listed in Table \ref{tab:lattice_const}, which are somewhat different
from those of structure $L$.  
The calculated Wyckoff positions (WPs) for the structure 
$L$ are given in Table \ref{tab:atompositions}, 
comparing with the experimental data, \cite{alonso97}
whereas the structure $R$ is a mirror image of $L$ about the $ac$-plane. 
The calculated WPs positions are  
extremely close to what was obtained experimentally.
However, the WPs of structure $L$ split by small amount 
from the WPs of the ($Pbam$) experimental structure, and
lower the structural symmetry to the long searched
\emph{Pb}2$_1$\emph{m} polar group. \cite{kagomiya03}
The relationships between the WPs of TbMn$_2$O$_5$ under $Pbam$ symmetry
and under the $Pb2_1m$ symmetry are given in
Table \ref{tab:spacegroup}.
For example, under the $Pbam$ symmetry, Mn$^{3+}$ has one WP
(notion h),with four equivalent sites. 
It splits into two WPs b(1) and b(2), 
each having two equivalent sites, when the symmetry lowers to $Pb2_ 1m$.

We artifically construct a high symmetry structure $H$
by symmetrizing structure $L$ and $R$ 
according to the \emph{Pbam} symmetry. \cite{wang07} 
The WPs of $H$ together with the atomic displacements from $H$ to $L$ 
are also given 
in Table \ref{tab:atompositions}, where
$\delta {\rm a}$, $\delta {\rm b}$ and $\delta {\rm c}$
denote the atomic displacements from the high symmetry positions
along the $a$, $b$ and $c$ axes respectively. 
The displacements along the $a$ and $c$ axes 
are of mirror symmetry, whereas the
displacements along the $b$ axis are not.
As seen in Table \ref{tab:atompositions}, all
cations move up and anions (except O$_1$) move down
from the positions of high symmetry structure $H$, 
resulting in polarization.
However, all the atomic displacements are extremely small,
usually are of the order of $\sim$ 10$^{-4}$ of the lattice
constants.
Therefore the low symmetry structure can not be directly
determined experimentally, and only the anomalies of 
the atomic displacement parameters (ADPs)
were observed. \cite{chapon04}
The largest atomic displacements come from Mn$^{3+}$,
and O$_3$, $\delta y$ $\sim$ 10$^{-3}$ of the lattice constants,
in agreement with the ADP results of Ref.\onlinecite{chapon04}.

\begin{widetext}
\begin{table*}
\begin{center}
\tabcolsep 2.5mm 
\caption{Comparison of the calculated and measured
atom positions of TbMn$_2$O$_5$. $L$ and $H$ are the ground state
structure and the high-symmetry structure respectively. 
$|\delta {\rm a}|$, $|\delta {\rm b}|$, and $|\delta \rm{c}|$
denote the atomic displacements from $H$ to $L$. 
The experimental values are taken from Ref. \onlinecite{alonso97}.}
\begin{tabular}{c ccc ccc crc ccc}
\hline \hline & \multicolumn{3}{c}{$L$ ($Pb2_1m$)}
&\multicolumn{3}{c}{$H$ ($Pbam$)} & \multicolumn{3}{c}{$H\rightarrow
L(10^{-4})$} &\multicolumn{3}{c}{Exp.
($Pbam$)} \\
atom &a &b & c &a &b  &c  & $|\delta {\rm a}|$ &$|\delta {\rm b}|$ &$|\delta
\rm{c}|$ &a &b &c\\
\hline
Tb$^{3+}_1$& 0.1410  & 0.1733 & 0       & 0.1407 & 0.1732 & 0      & 3.0 & 1.5  & 0        &  0.1399 &  0.1726  & 0      \\
Tb$^{3+}_2$& 0.6404  & 0.3270 & 0       &        &        &        & 3.0 & 1.5  & 0        &                             \\
Mn$^{4+}$  & 0.0001  & 0.5003 & 0.2558  & 0      & 0.5    & 0.2558 & 0.8 & 2.9  & 0        &  0      &  0.5     & 0.2618 \\
Mn$^{3+}_1$& 0.4012  & 0.3558 & 0.5     & 0.4014 & 0.3551 & 0.5    & 2.2 & 6.6  & 0        &  0.4120 &  0.3510  & 0.5    \\
Mn$^{3+}_2$& 0.9016  & 0.1456 & 0.5     &        &        &        & 2.2 & 6.6  & 0        &                             \\
O1         & 0.0008  & 0.0002 & 0.2709  & 0      & 0      & 0.2709 & 8.2 & 2.3  & 0        &  0      &  0       & 0.2710 \\
O2$_1$     & 0.1646  & 0.4480 & 0       & 0.1647 & 0.4481 & 0      & 1.2 & 1.2 & 0        &  0.1617 &  0.4463  & 0      \\
O2$_2$     & 0.6648  & 0.0517 & 0       &        &        &        & 1.2 & 1.2 & 0        &                             \\
O3$_1$     & 0.1560  & 0.4329 & 0.5     & 0.1565 & 0.4337 & 0.5    & 5.3 & 7.8 & 0        &  0.1528 &  0.4324  & 0.5    \\
O3$_2$     & 0.6571  & 0.0655 & 0.5     &        &        &        & 5.3 & 7.8 & 0        &                             \\
O4$_1$     & 0.3977  & 0.2077 & 0.2438  & 0.3968 & 0.2079 & 0.2430 & 8.8 & 2.2 & 8.5 &  0.3973 &  0.2062  & 0.2483 \\
O4$_2$     & 0.8959  & 0.2919 & 0.7579  &        &        &        & 8.8 & 2.2 & 8.5 &                             \\
\hline
\end{tabular}
\label{tab:atompositions}
\end{center}
\end{table*}
\end{widetext}

\begin{table}
\begin{center}
\tabcolsep 5mm \caption{The Wyckoff positions (WPs) 
for each inequivalent atom in TbMn$_2$O$_5$
under the space group $Pbam$ and and 
its subgroup $Pb2_1m$.}
\begin{tabular}{ccc}
\hline \hline
 Atoms & WP($Pbam$) & WP($Pb2_1m$)\\
\hline
Tb$^{3+}$ & 4g & 2a, 2a\\
Mn$^{4+}$ & 4f & 4c    \\
Mn$^{3+}$ & 4h & 2b, 2b\\
O1        & 4e & 4c    \\
O2        & 4g & 2a, 2a\\
O3        & 4h & 2b, 2b\\
O4        & 8i & 4c, 4c\\
\hline
\end{tabular}
\label{tab:spacegroup}
\end{center}
\end{table}

\begin{figure}
\centering
\includegraphics[width=3.2in]{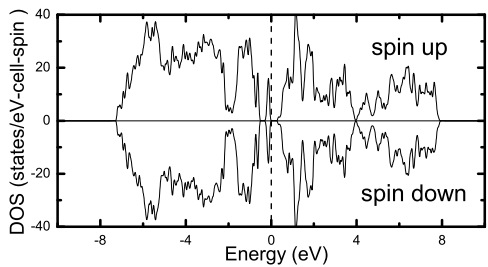}
\caption{The total density of sates (DOS) of TbMn$_2$O$_5$ for spin-up and
  spin-down respectively, 
calculated under structure $L$.
The dashed line represents the Fermi level.}
\label{fig:DOS}
\end{figure}

\begin{figure}
\centering
\includegraphics[width=3.5in]{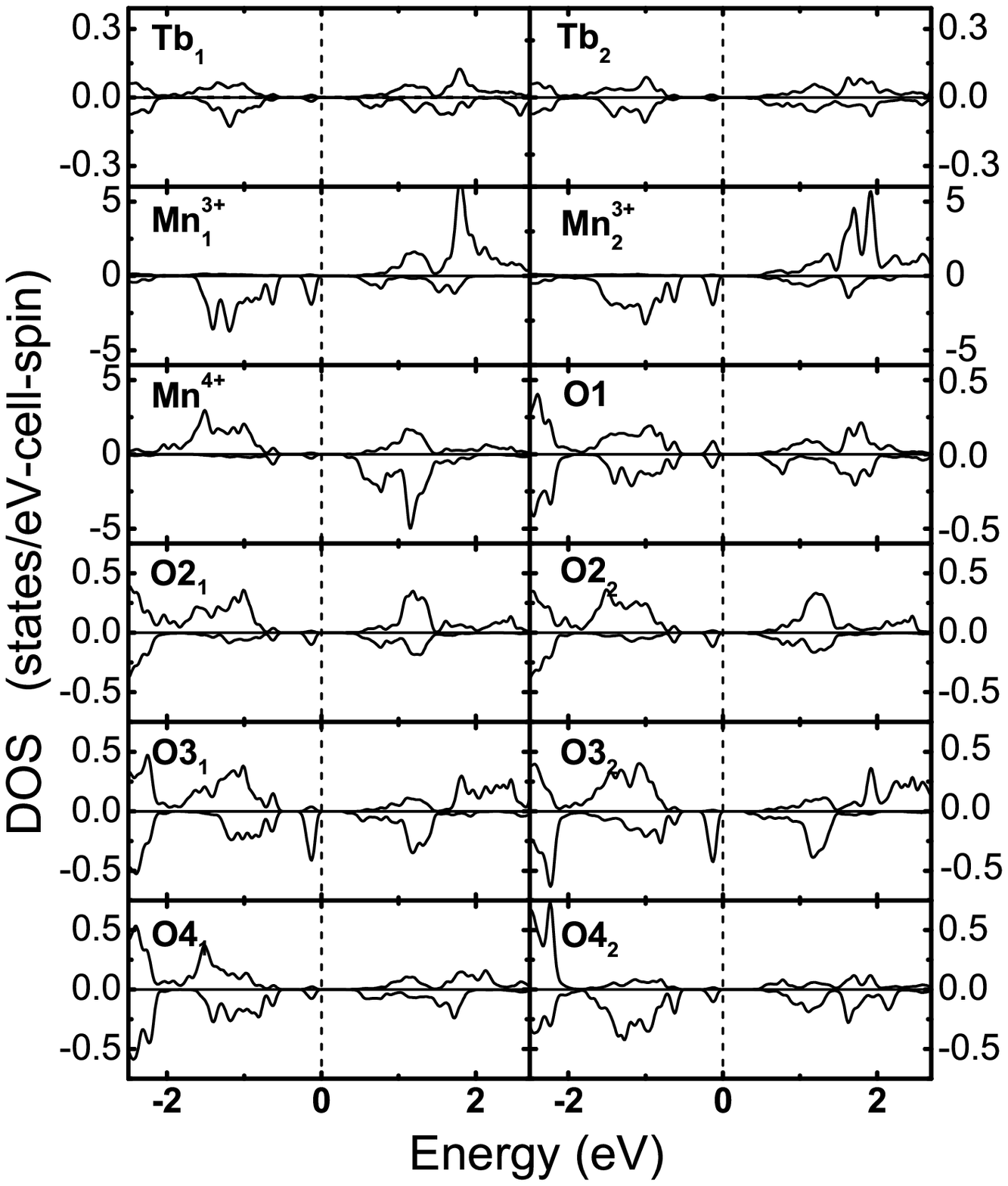}
\caption{Site-projected partial density of states (PDOS) of
Mn and Tb $d$ electrons, and O \emph{p} electrons for TbMn$_2$O$_5$. 
The upper panel and lower panel show the PDOS for
spin up and spin down respectively.
The PDOS of other states are
relatively small in the selected energy range, and therefore not shown. 
The dashed lines
represent the Fermi level. 
}
\label{fig:PDOS}
\end{figure}

\section{Electronic Structure} 
\label{sec:e_structure}

Figure \ref{fig:DOS} depicts the calculated the total densities 
of states (DOS) of structure $L$ . 
The DOS of spin-up and spin-down electrons are identical as expected 
for an AFM state. From the results, the band gap is estimated to
be $\sim$ 0.4 eV, confirming the experimental fact that TbMn$_2$O$_5$ is an
insulator. However, it is well known that GGA greatly 
underestimates the band gap, especially for the 3$d$ compounds, 
therefore the real band gap might be much larger.
We further calculated the site-projected partial DOS (PDOS) 
for Tb, Mn $d$ electrons
and O $p$ electrons in the energy range of -2.5 to 2.5 eV 
around the Fermi level, shown in Fig.\ref{fig:PDOS}. 
In structure $L$, the Tb, Mn$^{3+}$, O$_2$, O$_3$ and O$_4$ each has two 
non-equivalent sites, and the PDOS are shown separately.  
The PDOS for other
electrons are quite small in the selected energy range and therefore are not
shown. 
We see the DOS near the Fermi level is mainly from Mn 3$d$ and 
O 2$p$ orbitals . 

The local magnetic moments are estimated for Mn$^{3+}$ to be $\sim$ 
2.37 $\mu_B$, and for Mn$^{4+}$ to be $\sim$ 1.64 $\mu_B$, 
in a good agreement with the refined magnetic moments. \cite{chapon04}  
The ionic radii of 0.60 \AA\, and 0.66 \AA\, are used
for Mn$^{4+}$ and Mn$^{3+}$ respectively taken from the periodic table. 
The local magnetic moments of Mn$^{3+}$, and Mn$^{4+}$ at different
WPs are slightly different.
O ions also have small induced magnetic moments, due to their hybridization
with Mn ions. 
The calculated local moments of Tb are extremely small,
whereas 1.6 $\mu_B$ is assumed in order to obtain
a perfect fit of the neutron diffraction data in Ref. \onlinecite{chapon04}. 
The difference might come from that the $f$ electrons are not included
as valence electrons in the Tb pseudopotential.

We also calculated the DOS and PDOS of structure $H$ and 
find no obvious differences from structure $L$,
suggesting that the effects of the 
lattice structure distortion and the symmetry lowering
in structure $L$ to the electronic structure are small. 

\begin{table}
\tabcolsep 3mm \caption{The relationships of irreducible
representations between space group \emph{Pbam} and its subgroup
\emph{Pb}2$_1$\emph{m}.}
\begin{center}
\begin{tabular}{c ccc ccc ccc}
\hline \hline \mc{2}{c}{\emph{Pb}2$_1$\emph{m}} & \mc{2}{c}{$Pbam$}\\
irreps & Character & irreps & Character \\
\hline
 A$_1$ & IR and Raman & A$_g$    & Raman \\
       &              & B$_{2u}$ & IR    \\
 B$_1$ & IR and Raman & B$_{3g}$ & Raman \\
       &              & B$_{1u}$ & IR    \\
 B$_2$ & IR and Raman & B$_{1g}$ & Raman \\
       &              & B$_{3u}$ & IR    \\
 A$_2$ & Raman        & B$_{2g}$ & Raman \\
       &              & A$_{u}$  & Silent\\
\hline
\end{tabular}
\label{tab:irreps}
\end{center}
\end{table}

\section{Lattice dynamic properties}
\label{sec:phonon}

The symmetry lowering and ferroelectricity in the
traditional ferroelectrics, such as BaTiO$_3$, 
PbTiO$_3$, etc. are driven by the condensation
of zone center soft phonons.\cite{lines_book}
In TbMn$_2$O$_5$, the low symmetry ($Pb2_1m$) structure was not
observed {\it directly} in the experiments.   
However, a strong evidence that suggests the 
symmetry lowering is that one of the Raman-active phonons 
become also IR-active at low
temperature,\cite{aguilar06} 
which is forbidden by the higher \emph{Pbam} symmetry. 
To elucidate the origin of the symmetry lowering, 
we calculate the frequencies of all 
zone center phonons, using frozen phonon method, 
to exam if any soft modes exist in TbMn$_2$O$_5$ and the symmetry
properties of all phonons.
Furthermore, the validity of our theories can be examined by
comparing the calculated phonons with experiments.

\subsection{symmetry analysis}

We first carry out symmetry analyses and 
decompose the symmetry-adapted modes 
into irreducible representations.
The Hamiltonian remains invariant under a lattice translation of the 32-atom
primitive cell combined with spin reversal. Therefore the force-constant
matrix has the translational symmetry of the 32-atom primitive cell. \cite{he02}
We then perform the symmetry analyses on the 32-atom primitive
cell (instead of the 64-atom AFM cell used in the calculations), 
using the SMODES symmetry-analysis software package. \cite{smodes}

For the high symmetry structure (\emph{Pbam}),
all the 96 modes, including 3
acoustic modes, are decomposed into 8 irreducible representations
(irreps):
%
\begin{eqnarray}
\Gamma(Pbam)=&& 9B_{1u}\oplus15B_{2u}\oplus15B_{3u}\oplus13A_{g} \nonumber \\
&&\oplus13B_{1g}\oplus11B_{2g}\oplus11B_{3g}\oplus9A_{u}\,.
\label{eq:high-symm-modes}
\end{eqnarray}
%
Among them $B_{3u}$, $B_{2u}$ and $B_{1u}$ modes are IR active,
polarized along the \emph{a}, \emph{b} and \emph{c} axis, respectively,
whereas $B_{1g}$, $B_{2g}$, $B_{3g}$ and $A_{g}$ modes are Raman active and
$A_{u}$ modes are silent. All these irreps are one dimensional, i.e., no
degenerate phonons.
The three acoustic modes are each in one of the three IR-active representations.
As we see, the Raman- and IR-active modes do not couple under the $Pbam$ symmetry.

For the crystal structure processes
a \emph{Pb}2$_1$\emph{m} symmetry,
the 96 phonons can be decomposed into 4 irreps:
\begin{equation}
\Gamma(Pb2_1m) =28A_1\oplus20B_1\oplus28B_2\oplus20A_2 \, .
\label{eq:low-symm-modes}
\end{equation}
We found all modes are Raman active,
among them $A_1$, $B_1$ and $B_2$ are also IR active,
which are polarized along the $b$, $c$, and $a$ axis 
respectively. \cite{aguilar06} 
The relationships between the irreps of {\it Pbam} symmetry and its
subgroup {\it Pb}2$_1${\it m} symmetry are given in Table \ref{tab:irreps},
together with the symmetry character for each irrep.
The $A_1$ modes are coupled from the $A_{g}$
and $B_{2u}$ modes and the $A_2$ modes are composed of the $B_{2g}$ and $A_u$ modes.
The $B_1$ modes are the coupled $B_{3g}$ and $B_{1u}$ 
modes whereas the $B_2$ modes are the coupled $B_{1g}$  and $B_{3u}$
modes.

\begin{table*}
\tabcolsep 2mm 
\caption{Calculated phonon frequencies
$\omega$, effective charges $Z_{\lambda}^{*}$ and oscillator
strength $S_{\lambda}$ of $A_1$ modes, classified according
to their major symmetry characters and compared with the experimental data.
The IR-active phonons are extracted from Ref. \onlinecite{aguilar06} at
7K and Raman-active phonons are extracted from Ref. \onlinecite{mihailova05}.}
\begin{center}
\begin{tabular}{ccrcccc ccrcc}
\hline \hline \mc{7}{c}{$B_{2u}$(IR, $b$-polarized)} &
\mc{5}{c}{$A_g$
(Raman)} \\
\mc{2}{c}{$\omega$} & \mc{2}{c}{$Z_{\lambda}^{*}$} &
\mc{2}{c}{S$_{\lambda}$} & Raman\%&\mc{2}{c}{$\omega$} &
$Z_{\lambda}^{*}$ &
$S_{\lambda}$ & IR\%\\
GGA & Exp. & GGA & Exp. & GGA & Exp. & & GGA & Exp. & GGA & GGA &\\
\hline
100.7 & 97.2  & 0.09 & 0.17 & 0.11 & 0.42 & 0.08 & 110.1 &     & 0.020 & 0.005   & 0.02\\
158.0 & 168.9 & 0.28 & 0.31 & 0.44 & 0.46 & 0.12 & 136.9 &     &-0.003 & $\sim0$ & 0.17\\
162.8 & 171.9 & 0.31 & 0.25 & 0.49 & 0.30 & 0.04 & 221.6 & 215 & 0.058 & 0.009   & 0.94\\
224.8 & 222.2 & 0.30 & 0.20 & 0.24 & 0.11 & 1.85 & 235.1 & 221 & 0.066 & 0.011   & 1.38\\
267.3 & 256.8 & 0.25 & 0.28 & 0.12 & 0.17 & 0.36 & 312.5 & 334 & 0.125 & 0.022   & 0.97\\
316.7 & 333.4 & 0.78 & 0.37 & 0.82 & 0.17 & 1.02 & 340.2 & 350 & 0.094 & 0.011   & 2.96\\
351.3 & 386   & 0.28 & 0.15 & 0.09 & 0.02 & 2.94 & 405.6 & 412 &-0.028 & $\sim0$ & 0.24\\
412.5 & 422.3 & 0.39 & 0.60 & 0.13 & 0.28 & 0.23 & 445.1 &     & 0.423 & 0.126   & 2.24\\
439.5 & 453.2 & 2.58 & 2.25 & 4.81 & 3.43 & 2.65 & 489.2 & 500 & 0.151 & 0.013   & 6.51\\
471.0 & 481.8 & 1.40 & 2.18 & 1.23 & 2.86 & 5.89 & 529.2 & 537 &-0.104 & 0.005   & 6.16\\
533.5 & 538.2 & 0.52 & 0.72 & 0.27 & 0.25 & 5.95 & 612.0 & 621 &-0.002 & $\sim$0 & 3.28 \\
549.3 & 567.3 & 0.50 & 1.09 & 0.12 & 0.52 & 1.33 & 613.5 & 631 & 0.030 & $\sim$0 & 9.59\\
625.0 & 636.6 & 1.01 & 0.88 & 0.36 & 0.27 &11.20 & 673.6 & 693
\footnote{IR-active and 703 $cm^{-1}$ measured by Aguilar\cite{aguilar06}} &-0.107 & 0.004   & 8.04\\
667.2 & 688.3 & 0.17 & 0.10 & 0.01 & 0.003& 8.83 &       &     &       &         & \\
\hline
\end{tabular}
\label{tab:A1phonon}
\end{center}
\end{table*}

\begin{table*}
\tabcolsep 2mm 
\caption{Calculated phonon frequencies
$\omega$, effective charges $Z_{\lambda}^{*}$ and oscillator
strength $S_{\lambda}$ of $B_2$ modes, classified according
to their major symmetry characters and compared with the experimental data.
The IR-active phonons are extracted from Ref. \onlinecite{aguilar06} at
7K and Raman-active phonons extracted from Ref. \onlinecite{mihailova05}.}
\begin{center}
\begin{tabular}{ccccccc ccccc}
\hline \hline \mc{7}{c}{$B_{3u}$(IR, $a$-polarized)}&
\mc{5}{c}{$B_{1g}$(Raman)} \\
\mc{2}{c}{$\omega$} & \mc{2}{c}{$Z_{\lambda}^{*}$} &
\mc{2}{c}{S$_{\lambda}$} & Raman\%& \mc{2}{c}{$\omega$} &
$Z_{\lambda}^{*}$ &
$S_{\lambda}$ & IR\% \\
GGA & Exp. & GGA & Exp. & GGA & Exp. & & GGA & Exp. & GGA & GGA & \\
\hline
111.8 & 111.9 & 0.02 & 0.23 & 0.004 & 0.59 & 0.02 & 148.2 & 148 & 0.007 & $\sim$0 & 0.54\\
152.2 & 157.5 & 0.17 & 0.37 & 0.18  & 0.81 & 0.50 & 172.9 & 172 & 0.006 & $\sim$0 & 0.13\\
159.8 & 164.2 & 0.76 & 0.57 & 3.11  & 1.68 & 0.12 & 222.1 & 208 & 0.054 & 0.008   & 10.5\\
217.4 & 218.5 & 0.05 & 0.32 & 0.01  & 0.30 & 10.99& 243.0 & 237 & 0.067 & 0.011   & 0.97\\
264.5 & 254.8 & 0.30 & 0.94 & 0.18  & 1.88 & 0.62 & 300.7 & 326 & 0.007 & $\sim$0 & 0.18\\
317.0 & 333.1 & 0.03 & 0.27 & 0.002 & 0.09 & 0.15 & 373.0 &     & 0.048 & 0.002   & 9.50\\
353.1 & 364.9 & 1.04 & 1.39 & 1.21  & 2.02 & 1.25 & 410.4 & 416 & 0.019 & $\sim$0 & 0.31\\
379.1 & 397.6 & 0.56 & 0.66 & 0.30  & 0.38 & 8.44 & 470.0 & 485 & 0.360 & 0.081   & 36.8\\
465.2 &       & 0.21 &      & 0.03  &      & 1.20 & 518.3 &     & 0.115 & 0.007   & 2.82\\
474.4 & 494.8 & 0.24 & 0.89 & 0.03  & 0.45 & 37.76& 545.8 & 538 & 0.054 & 0.001   & 0.58\\
509.2 &       & 1.18 &      & 0.75  &      & 0.59 & 587.1 &     & 0.178 & 0.013   & 1.91\\
595.4 & 613.5 & 0.90 & 1.38 & 0.32  & 0.71 & 1.28 & 643.8 & 673 & 0.445 & 0.067   & 5.35\\
617.1 & 627.5 & 1.23 & 0.81 & 0.55  & 0.23 & 4.57 & 678.7 & 697 & 0.027 & $\sim$0 & 10.0\\
672.7 & 704.2 & 0.33 & 0.42 & 0.03  & 0.05 & 12.12&       &     &       &         &\\
\hline
\end{tabular}
\label{tab:B2phonon}
\end{center}
\end{table*}

\subsection{Zone Center Optical Phonons}
\label{sec:fozen phonon}

The zone-center phonons frequencies 
are calculated via a frozen phonon technique. \cite{he02}
For each irrep $\Gamma$ in 
Eq. (\ref{eq:high-symm-modes}), (\ref{eq:low-symm-modes}), 
we construct an $N_{\Gamma}\times N_{\Gamma}$ dynamical matrix 
from a series of
frozen-phonon calculations in which the structure is distorted
according to each of the $N_{\Gamma}$ symmetry modes in this
irrep. For example, $N_{\Gamma}$=28 for the A$_1$ modes
in Eq. (\ref{eq:low-symm-modes}). 
All calculations were performed in
the 64-atom AFM cell respect to the correct AFM
reference ground state.
After calculating the residual Hellmann-Feynman forces 
we displace the ions according to
the symmetry coordinate for each mode by 0.1\% of lattice constant
and calculate the forces $F_i^{\alpha}$ on ion $i$ in
Cartesian direction $\alpha$ due to displacements $u_j^{\beta}$
of ion $j$ in direction $\beta$. The force constant matrix
\begin{equation}
\Phi^{\alpha \beta}_{ij} 
=-{\partial F^{\alpha}_i \over \partial u^{\beta}_j}\, ,
\end{equation}
is then obtained by finite differences to the residual forces in 
the equilibrium structure to eliminate the numerical
errors due to the residual forces.
The normal modes ${\bf u}_\lambda$ and their frequencies
$\omega_{\lambda}$ are then obtained through solution of the
eigenvalue equation
\begin{equation}
{\bf \Phi}\cdot{\bf u}_\lambda =\omega_\lambda^2 \,
{\bf M}\cdot{\bf u}_\lambda.
\end{equation}
Here $M_{ij}^{\alpha\beta}=(m_i/m_0)\,\delta_{ij}\,\delta_{\alpha\beta}$
is the dimensionless diagonal mass matrix, where $m_i$ is the mass
of atom $i$ and $m_0$ is a reference mass chosen here to be
1\,amu, and the eigenvectors are normalized
according to ${\bf u}_\mu \cdot {\bf M} \cdot {\bf u}_\nu =
\delta_{\mu\nu}$.

The phonon frequencies calculated from
high structure $H$ are extremely close to those calculated from low symmetry
structure $L$.
No soft mode is found in both high and low
symmetry structures, excluding the possibility of 
ferroelectricity driven by the soft modes.
We therefore show only the phonon frequencies for the ground state
structure $L$ only.
The calculated $A_1$, $B_2$, $B_1$ phonon frequencies are given 
in Table \ref{tab:A1phonon}, Table \ref{tab:B2phonon},  and
Table \ref{tab:B1phonon} respectively, whereas the $A_2$
phonons are given in Table \ref{tab:A2phonon}.
To make a good contact with experiments, we divide
the phonons of each irrep into two
presentations according to their major symmetry characters 
(see Table \ref{tab:irreps}).
We found that for most of the modes 
the coupling between the major and minor characters
(IR and Raman or Raman and silent)  
are extremely small, 
and such classification is unambiguous.
All calculated phonon frequencies are
in excellent agreement with the experiments especially
for the low-frequency phonons. The errors are
less than 5\%. 

In order to check the influences on phonons from different magnetic
orderings and magnetic-ordering-induced lattice modulations,
we also calculated the phonon frequencies for the FM structure,  we find
the differences are also small. 

In the following sections, we compare the 
calculated IR and Raman spectra with the experiments in details.

\begin{table*}
\tabcolsep 1.5mm 
\caption{Calculated phonon frequencies
$\omega$, effective charges $Z_{\lambda}^{*}$ and oscillator
strength $S_{\lambda}$ of $B_1$ modes, classified according
to their major symmetry characters.
The Raman-active phonons
are extracted from Ref. \onlinecite{mihailova05}.}
\begin{center}
\begin{tabular}{cccc ccccc}
\hline \hline \mc{4}{c}{$B_{1u}$(IR, $c$-polarized)}&
\mc{5}{c}{$B_{3g}$(Raman)} \\
$\omega$ & $Z_{\lambda}^{*}$ & S$_{\lambda}$ & Raman\%&
\mc{2}{c}{$\omega$} & $Z_{\lambda}^{*}$ &
$S_{\lambda}$ & IR\% \\
GGA & GGA & GGA & & GGA & Exp. & GGA & GGA & \\
\hline
142.2 & 0.49 & 1.65 & $\sim$0 & 102.6 &     & 0.005 & $\sim$0 & 0.004\\
249.5 & 1.01 & 2.27 & 0.62    & 192.1 & 195 & 0.007 & $\sim$0 & 0.03 \\
300.2 & 0.18 & 0.05 & 3.28    & 241.6 & 244 & 0.100 & 0.024   & 0.41 \\
407.4 & 1.47 & 1.828 & 24.58  & 265.5 &     & 0.040 & 0.003   & 0.69 \\
433.9 & 0.84 & 0.53 & 27.42   & 289.0 & 299 & 0.107 & 0.019   & 1.50 \\
495.8 & 1.20 & 0.82 & 36.47   & 325.2 &     & 0.098 & 0.013   & 1.27 \\
527.2 & 0.39 & 0.08 & 2.80    & 426.4 &      & 2.25   & 3.89   & 48.8 \\
605.1 & 0.74 & 0.21 & 0.56    & 447.2 & 442  & 0.116 & 0.009   & 0.45 \\
      &      &      &         & 470.0 &      & 0.124 & 0.010   & 5.92 \\
      &      &      &         & 505.8 & 540  & 0.909 & 0.450   & 34.5 \\
      &      &      &         & 553.9 & 577  & 0.012 & $\sim$0 & 2.07 \\
\hline
\end{tabular}
\label{tab:B1phonon}
\end{center}
\end{table*}

\subsubsection{IR spectra and dielectric properties}
\label{sec:ir}

Aguilar {\it et al.}  measured the frequencies of IR-active
phonon polarized along the $b$ and $a$ axes. \cite{aguilar06}
These phonon modes are compared with the calculated modes in 
Table \ref{tab:A1phonon} and Table \ref{tab:B2phonon}, 
respectively.

For the IR phonons, we can further calculate their 
contribution to the dielectric constant.
The dielectric function tensor in the frequency range near the phonon frequencies
could be written as sum
\begin{equation}
\epsilon_{\alpha\beta}(\omega)=(\epsilon_{\infty})_{\alpha\beta}+ 
(\epsilon_{\rm ph})_{\alpha\beta} (\omega)\; 
\end{equation}
of electronic contribution and lattice contribution, where 
$\alpha$, $\beta=\hat{\bf a}, \hat{\bf b}, \hat{\bf c}$ 
are the axis indices.
For TbMn$_2$O$_5$, $(\epsilon_{\infty})_{aa}$ and
$(\epsilon_{\infty})_{bb}$ were measured to be 5.31 and 6.82
respectively.\cite{aguilar06}
In TbMn$_2$O$_5$, the phonons are polarized along the $a,b,c$ axis, 
and $\epsilon_{\alpha\beta}(\omega)$ is diagonal. We could therefore ignore
the subscripts $\alpha$, $\beta$, but only give the polarization direction.
The dielectric contribution from the $\alpha$-polarized 
IR-active phonons is then
\begin{equation}
\epsilon_{\rm ph}^{(\alpha)}(\omega)= \Omega_0^2 \sum_{\lambda}
{Z_{\lambda}^{*2} \over \omega_{\lambda}^{2}-\omega^{2}}
\; ,
\end{equation}
where  $\omega_\lambda$ and $Z^*_{\lambda}$ are the
mode frequencies and mode dynamical charges respectively,
and ${\Omega_0}^2= 4\pi e^2/m_0 V$ is a characteristic 
frequency having the interpretation of a plasma frequency 
of a gas of objects of mass 
$m_0$=1\,amu, charge $e$, and density $V^{-1}$
($V$ is the 32-atom primitive cell volume). \cite{he02}
The sum is restricted to the $\alpha$-polarized phonons. 
The mode dynamical charges are defined as
\begin{equation}
Z_{\lambda}^{*}=V {\partial P_{\alpha} \over \partial
u_{\lambda}}
\end{equation}
in which $P_{\alpha}$ is the polarization in the $\alpha$ direction
due to a small frozen-ion amplitude
of the $\lambda$-th phonon modes. \cite{he02}
In practice we compute the mode dynamical charges by finite
differences method following Ref. \onlinecite{he02}.
Once we have the mode effective charges, we calculate the oscillator
strengths of each IR-active modes
$S_{\lambda}= \Omega_0^{2}Z_{\lambda}^{*2}
/\omega_{\lambda}^2$,
and the total lattice
contribution to the static ($\omega\rightarrow 0$) dielectric constant
$\epsilon_{\rm ph}(0) = \Omega_0^2  \sum_\lambda Z_{\lambda}^{*2}
/ \omega_{\lambda}^2 =\sum_\lambda S_\lambda$.

The results of the Born effective charges and oscillator strengths for the
IR-active modes are also listed in Table \ref{tab:A1phonon}, \ref{tab:B2phonon}
and \ref{tab:B1phonon}, for 
the $A_1$ ($b$-polarized), $B_2$($a$-polarized) 
and $B_1$ ($c$-polarized) phonons respectively. 

Let us first look at the $A_1$ ($b$-polarized) modes presented 
in Table \ref{tab:A1phonon}. There are 28 $A_1$ modes in total, and
14 (plus one acoustic mode) of them that are mainly $B_{2u}$ IR modes.
The calculated phonon frequencies and oscillator strengths are in very good
agreement with experiments. The calculated total oscillator strength 
is about 9.24 compared with the measures value 9.12.  
The rest 13 modes
are mainly $A_g$ Raman modes. Under the low $Pb2_1m$ symmetry, 
these modes also acquire some small oscillator strengths. 
More importantly, the 693 cm$^{-1}$ $A_g$ Raman mode
(703 cm$^{-1}$ measured from IR spectra \cite{aguilar06}), 
that was found also IR active with oscillator
strength $S_{\lambda}$=0.001 in the experiment, \cite{aguilar06}
is well reproduced in the calculations,
with oscillator strength  $S_{\lambda}$=0.004, 
therefore, confirming that
the ground state structure is indeed of \emph{Pb}2$_1$\emph{m}
symmetry. \cite{wang07}
The oscillator strengths
of other $A_g$ modes might be 
covered by the adjacent $B_{2u}$ modes, and therefore 
are not observed experimentally.
To see more clearly between the coupling of the IR and Raman modes,
we also list the percentage of the characters in the Table.
We see the coupling between IR and Raman modes is fairly small.

The results of 28 $B_2$ phonons are given in Table \ref{tab:B2phonon},
classified into two presentation according to their major symmetry characters,
including 14 IR ($B_{3u}$) modes, 13 ($B_{1g}$) Raman modes and one acoustic mode. 
Two of the $B_{3u}$ modes were missing in the experiments, \cite{aguilar06}
possibly due to their small oscillator strengths. 
Here we tentatively assign them as the 465.2 and 509.2 cm$^{-1}$ modes.
Again the phonon frequencies agree very well with the experiments. However, the
IR oscillator strengths are in less agreement with the experimental values. 
Some modes have large discrepancy in the oscillator strengths
compared to the experiments,
e.g, the 264 cm$^{-1}$ mode. Nevertheless, the calculated 
total oscillator strength for $a$-polarized
phonons is about 6.87, in a reasonable agreement with the experiments value
about 9.19. \cite{aguilar06}
Similar to the $A_g$ modes, the $B_{1g}$ Raman modes also have small IR oscillator
strengths due to the $Pb2_1m$ symmetry. For most modes, the coupling between
the IR and Raman characters are small, except two modes: the 474 cm$^{-1}$
mode in the $B_{3u}$ presentation and the 470 mode in the $B_{1g}$
presentation. This is probably because they are {\it accidentally} degenerate
in phonon frequencies, leading to large character mixing.  

The phonon frequencies and oscillator strengths of the $c$-axis polarized
$B_1$ phonons are given in Table \ref{tab:B1phonon},
including 8 (plus one acoustic mode) $B_{1u}$ IR modes and 11 $B_{3g}$ Raman
modes.
Interestingly, there are some modes that have 
very large IR and Raman modes mixing in this
representation. For example, the 426 $cm^{-1}$ Raman modes have about 49\% IR
character and very large oscillator strength of $S_{\lambda}$=3.89.
However, this large Raman-IR mixing might come from that
this modes accidentally degenerate with the
407 and 433  $cm^{-1}$ IR modes in the numerical calculations, 
and might not be the case in the real system.   
Unfortunately, we can not find suitable experimental 
IR spectra to compare with for the $B_1$ modes.  
We hope future experiment can clarify this problem.

\begin{table}
\tabcolsep 1.5mm 
\caption{Calculated phonon frequencies
$\omega$ of $A_2$ modes, classified according to their major
symmetry characters and compared with the experimental data. 
The Raman-active phonons are extracted
from Ref. \onlinecite{mihailova05}.}
\begin{center}
\begin{tabular}{ccccc}
\hline \hline
\mc{2}{c}{$A_u$ (Silent)} &\mc{3}{c}{$B_{2g}$ (Raman)}\\
$\omega$ & Raman\% & \mc{2}{c}{$\omega$} & Raman\% \\
GGA & & GGA  & Exp. & \\
\hline
110.2 & 0.04 &  95.9 &     & 99.9 \\
131.4 & 0.09 & 210.6 & 214 & 99.4 \\
226.6 & 6.18 & 231.1 & 232 & 99.9 \\
293.9 & 40.7 & 274.2 &     & 99.4 \\
399.4 & 12.6 & 288.9 & 301 & 60.3 \\
425.0 & 3.00 & 337.5 &     & 99.3 \\
497.7 & 5.05 & 441.1 &     & 89.1 \\
529.4 & 1.96 & 448.7 & 455 & 96.9 \\
611.5 & 5.65 & 469.6 & 470 & 92.9 \\
      &      & 478.6 & 505 & 99.3 \\
      &      & 562.6 &     & 99.1 \\
\hline
\end{tabular}
\label{tab:A2phonon}
\end{center}
\end{table}

\subsubsection{Raman spectra}
\label{sec:raman}

The temperature-dependent Raman spectra of HoMn$_2$O$_5$ and TbMn$_2$O$_5$
were measured by Mihailova et al. \cite{mihailova05} 
We compare the calculated and measured Raman spectra of TbMn$_2$O$_5$ in
Table \ref{tab:B2phonon}, \ref{tab:A1phonon}, \ref{tab:B1phonon} and
\ref{tab:A2phonon} for the $B_{1g}$, 
$A_g$, $B_{3g}$ and $B_{2g}$ modes respectively.
The $A_u$ silent modes are also given in Table \ref{tab:A2phonon}, for they
also Raman-active in structure $L$.
Some Raman modes were missing in the experiments 
probably because the missing modes are of very low intensity or out of
the spectral range ($\omega<100 cm^{-1}$) of the experimental setup. 
\cite{mihailova05}
Once again, the calculated phonon frequencies are in very good agreement with
the experimental values.

So far, no IR-active modes has been observed in the Raman spectra of
TbMn$_2$O$_5$ .
However, there is an evidence that one of the IR modes become also Raman-active in 
HoMn$_2$O$_5$. \cite{mihailova05}
Three low frequency $A_g$ Raman modes were observed at 10 K in  HoMn$_2$O$_5$, 
\cite{mihailova05} and they are the 217
cm$^{-1}$, 219 cm$^{-1}$ and 226 cm$^{-1}$ modes, whereas
the 219 cm$^{-1}$ mode disappears at room temperature.
Since the phonons of frequencies at this range are mainly
due to the vibrations of Mn and O, and less related to the $R$ atoms,
one could therefore expect there are also 
three Raman modes in this frequencies range for TbMn$_2$O$_5$.
However, only two $A_g$ modes (215 cm$^{-1}$ and 221 cm${-1}$ modes) 
were observed (nevertheless at $T$= 300 K) in this frequency range in
TbMn$_2$O$_5$, consistent with our first-principles calculations
for the  $A_g$ modes.
We therefore propose that the 219 cm$^{-1}$ mode in HoMn$_2$O$_5$
is actually an IR-active phonon that become also Raman-active at low temperature.

\section{Discussions on the microscopic origin of the electric polarizations
  and giant magnetoelectric coupling}
\label{sec:polarization}

We have now firmly established that the ground
state structure of TbMn$_2$O$_5$
($L$ or $R$) is indeed of symmetry $Pb2_1m$,
allowing spontaneous polarization along the $b$-axis.
We have calculated \cite{wang07} the spontaneous polarization of structure $L$
($R$) via Berry phase technique. \cite{king-smith93}
The {\it intrinsic} polarization in this material is
calculated to be 1187 $nC\cdot$cm$^{-2}$. 
\cite{wang07}
This value is an order of magnitude smaller than 
that of the traditional ferroelectrics \cite{lines_book} (e.g, for
BaTiO$_3$ $\sim 25 \mu C\cdot$cm$^{-2}$), 
yet it is about 30 times larger than the {\it currently} 
measured value \cite{hur04} ($\sim$ 40 $nC\cdot$cm$^{-2}$ ) 
for this compound.
The reasons for the discrepancy between the
calculations and experiments might 
because of the approximations used in calculations.
For example, we approximate the magnetic propagation vector 
$k_z$=0.32 by zero. We also
ignore the spin-orbit coupling and assuming collinear spins. Without these
approximations, the polarization might be reduced. 
On the other hand, the experiment \cite{hur04} measured
polycrystalline samples,
in which the grains polarize along different directions cancel each other,
therefore might significantly underestimate the {\it intrinsic} polarization.
We believe a high quality single crystal sample should greatly
enhance  the measured electric polarization.

To further elucidate the origin of the polarization,
we calculate the spontaneous polarization for the high symmetry
structure $H$ and get ${\bf P}$= 228 $nC\cdot$cm$^{-2}$.
It might be surprising to see a crystal that possesses inversion symmetry would
develop spontaneous polarization at the first sight. 
However, to discuss the symmetry of a crystal with a magnetic ordering,
the pure point groups may not be enough.  
One has to resort to the magnetic groups to take account of the spin components. 
Based on the symmetry considerations, 
there are three possibilities (but not exclusive)
to develop spontaneous polarization:  

(i) The crystal structure itself does not have inversion symmetry. This is the
most familiar case to us, as it is seen in the traditional ferroelectrics. 

(ii) The crystal structure itself has the $R^{-1}$ symmetry, 
but the magnetic structure does not have the $R^{-1}$ symmetry {\it and} 
the combination of the $R^{-1}$ and time-reversal ($T^{-1}$) symmetry, 
$(RT)^{-1}$. 
If the magnetic structure of a system does not have the $R^{-1}$ symmetry, but
still has the $(RT)^{-1}$ symmetry, each spin channel may have finite electric
polarization. However, the polarizations from spin up and spin down channels 
cancel each other, leading to zero total electric polarization.

(iii) Even both the crystal and magnetic structure have the inversion 
symmetry,  but they do not have the common inversion center, 
as proposed by Betouras {\it et al.}. \cite{betouras07}

In the later two cases,  the ``electronic'' inversion symmetry is
broken because of the magnetic ordering. 
The electronic symmetry breaking will further couple to the 
lattice and lead to lattice distortion. 
It was shown \cite{wang07} that in TbMn$_2$O$_5$,
the SC $g$ does not have the $R^{-1}$ and the $(RT)^{-1}$ 
symmetry (case ii), because the special spin configurations 
of the Mn$^{3+}$-Mn$^{4+}$-Mn$^{3+}$ chains along the $b$ axis.   
Therefore, when holding atoms fixed at the centrosymmetric
structure, turning on the magnetic order does two things: it generates
a purely electronic polarization of 228 $nC\cdot$cm$^{-2}$, and
it also applies forces to the atoms.  
These forces give rise to atomic displacements that yield
an additional 959 $nC\cdot$cm$^{-2}$. 

The above discussion is based solely on the symmetry argument.
We now discuss the microscopic mechanism of coupling between the 
magnetic order and lattice.
The microscopic origin of the ME coupling may come from \cite{fiebig05}

(i)  Symmetric superexchange interactions that of the 
{\it scalar} field type: 
${\bf S_n} \cdot{\bf S_{n+1}}$. \\
(ii) Antisymmetric superexchange interactions that of the
{\it vector} field type:
${\bf S_n} \times {\bf S_{n+1}}$.\\ 
The antisymmetric superexchange interaction (also 
known as Dzyaloshinskii-Moriya interaction) originate
from the spin-orbit interaction. 
For the antisymmetric superexchange mechanism, 
nonocollinearity is essential for the ME coupling.
Experimentally, it was shown that in TbMn$_2$O$_5$ the largest electric
polarization is associated with a commensurate magnetic phase that is almost
collinear, \cite{chapon04, chapon06} therefore, in this phase, the spin-orbit
contribution to the ME coupling is small. An alternative mechanism must be
considered to account for the ME effects.
In the present calculations, we ignore the spin-orbit coupling and assume 
collinear spin (${\bf S}_i\parallel{\bf S}_j$),
therefore the symmetry breaking and ME coupling is merely due to the 
symmetric superexchange ${\bf S_n} \cdot{\bf S_{n+1}}$.

We now derive an effective model to describe the microscopic mechanism of the
ME coupling in TbMn$_2$O$_5$, 
based on a Heisenberg model assuming NN interactions. 
The total energy of the system around the high-symmetry structure 
$H$ can be written as,
\begin{equation}
E(\{u_{\lambda} \})=E_0+{1 \over 2}\sum_{\lambda}m_{\lambda}\omega_{\lambda}^{2}u_{\lambda}^{2}
-\sum_{ij}J_{ij}(\{u_{\lambda}\}){\bf S}_i\cdot {\bf S}_j \, ,
\end{equation}
where $u_{\lambda}$ is the $\lambda$-th zone-center phonon,
and ${\bf S}_i$ is the magnetic moment of the $i$-th atom. Here, we consider
only the magnetic moments of the Mn$^{3+}$ and Mn$^{4+}$ ions.
$E_0$ is the total energy of structure $H$ without spin-spin interactions. 
Since the magnetic-ordering-induced lattice distortion
from structure $H$ is small, we could
expand the exchange interaction $J_{ij}(\{u_{\lambda}\}) $ 
up to the second order of $u_{\lambda}$, i.e.,
\begin{widetext}
\begin{equation}
E(\{u_{\lambda} \}) =(E_0-\sum_{ij}J_{ij}(0) {\bf S}_i\cdot {\bf S}_j)
-\sum_{ij}\sum_{\lambda} {\partial J_{ij} \over \partial
u_{\lambda}} u_{\lambda} {\bf S}_i\cdot {\bf S}_j \\
+({1 \over 2}\sum_{\lambda} m_{\lambda} \omega_{\lambda}^{2}u_{\lambda}^{2}
-\sum_{ij}\sum_{\lambda\rho}{\partial^{2}J_{ij} \over \partial u_{\lambda}
\partial u_{\rho}}u_{\lambda}u_{\rho}{\bf S}_i\cdot {\bf S}_j)\, .
\end{equation}
\end{widetext}
The above three terms play different roles in the multiferroics.

(i) The zeroth order term in $u_{\lambda}$,
\begin{equation}
E_0-\sum_{ij}J_{ij}(0) {\bf S}_i \cdot {\bf S}_j\, ,
\end{equation}
determines the ground state spin configurations which is further discussed in 
Appendix \ref{sec:appendix}. \\

(ii) The second term in Eq. (\ref{eq:heff}) that is linear in $u_{\lambda}$, 
\begin{equation}
-\sum_{ij}\sum_{\lambda}{\partial J_{ij} \over \partial
u_{\lambda}}u_{\lambda}{\bf S}_i\cdot{\bf S}_j \, ,
\end{equation}
provides the driven forces for the
lattice distortion from high symmetry structure $H$.
The force on the $\lambda$-th phonon modes is
\begin{equation}
F_{\lambda}=-{\partial E \over \partial
u_{\lambda}} \vert_{u_{\lambda}=0}
=\sum_{ij}\frac{\partial J_{ij}}{\partial
u_{\lambda}}\textbf{S}_i\cdot\textbf{S}_j \, ,
\label{eq:force}
\end{equation}
which is nonzero, provided 
$\{ {\bf S}_i \}$ does not have the $R^{-1}$ and 
$(RT)^{-1}$ symmetry.

In Ref. \onlinecite{wang07}, we illustrated the lattice distortion
from the high symmetry structure using a spin chain model.
For SC $g$, inside the Mn$^{3+}$-Mn$^{4+}$-Mn$^{3+}$ segments along the
$b$-axis, Mn$^{4+}$ always have the same spin with the
upper Mn$^{3+}$ but opposite spin with the lower Mn$^{3+}$, whereas for
$g'$, the opposite is true.
Mn$^{4+}$ could therefore move closer to the Mn$^{3+}$ with same spin 
to lower the exchange energy.
According to Eq. (\ref{eq:force}), 
we have $F_{\lambda}(g)=-F_{\lambda}(g')$.
The high symmetry structure $H$ then spontaneous relax to structure $L$ or $R$
according to its SC.
The lattice distortion from high symmetry structure can be estimated as,
\begin{equation}
\delta\! u \sim {\partial J \over \partial u} /\bar{\omega}^2 \, ,
\end{equation}
where $\bar{\omega}$ is the {\it weight} averaged phonon frequency. 
We could therefore expect that a
material with larger ${\partial J / \partial u}$ and softer phonon frequencies
would have larger ME effects, 
provided the symmetry requirements are satisfied.

(iii) The quadratic term in $u_{\lambda}$, 
\begin{equation}
{1 \over 2}\sum_{\lambda} m_{\lambda}\omega_{\lambda}^{2}u_{\lambda}^{2}
-\sum_{ij}\sum_{\lambda\rho}\frac{\partial^{2}J_{ij}}{\partial u_{\lambda}
\partial u_{\rho}}u_{\lambda}u_{\rho} {\bf S}_i \cdot {\bf S}_j
\label{eq:heff}
\end{equation}
describes the coupling between phonons and magnons, which renormalizes
the phonon frequencies. For example, in ZnCr$_2$O$_4$,
\cite{sushkov05,fennie06} a
 large splitting between IR active phonons along 
the $x$ and $z$ axis was observed due
 to the phonon-magnon coupling.
The phonon-magnon coupling might be responsible to the
``step'' in dielectric constant at 24K in TbMn$_2$O$_5$ . 
Detailed investigation of the phonon-magnon coupling in TbMn$_2$O$_5$ 
is out of the scope of the present work,
and we leave it for future investigations.

\section{summary}
\label{sec:summary}

We have investigated the ground state structural and electronic properties
of multiferroic TbMn$_2$O$_5$ 
using density functional theory within GGA approximations
to elucidate the microscopic origin of the ferroelectrcity and giant
magnetoelectric coupling.
We use the collinear spin approximation and ignore spin-orbit coupling.
The calculated lattice and electronic structures agree well with 
the known experiments. 
We confirm that the ground state structure
of TbMn$_2$O$_5$ is of space group $Pb2_1m$, allowing polarizations.
The spontaneous electric polarization is calculated 
to be $1187$ $nC\cdot$cm$^{-2}$ along the $b$ axis.
The calculated zone-center optical phonons frequencies 
and the oscillator strengths of IR phonons
agree very well with the experimental values.
Unlike the traditional ferroelectrics, no soft phonons is found in
TbMn$_2$O$_5$. 
We derive an effective Hamiltonian to explain the magnetically-induced
ferroelectricity, in which the spin-lattice coupling is of 
symmetric superexchange interaction type.
Our results strongly suggest that the ferroelectricity in TbMn$_2$O$_5$ is
driven by the magnetic ordering that breaks the the inversion symmetry
without invoking the spin-orbit coupling.


\acknowledgments

L.H. would like to thank D. Vanderbilt for valuable suggestions. This work
was supported by the Chinese National Fundamental
Research Program 2006CB921900, 
the Innovation funds and ``Hundreds of Talents''
program from Chinese Academy of Sciences.

\begin{table}
\begin{center}
\tabcolsep 1.2mm \caption{
The DFT-GGA calculated total energies of spin configurations (shown in 
Fig. \ref{fig:spinconfigurations}) under the 
high symmetry structure $H$, where $E_3=-2J_3\,S_3\cdot S_4$,
$E_4=-2J_4\,S_3\cdot S_4$,  
$E_5=-2J_3\,S_3\cdot S_3$. The exchange
integrals $J_3$, $J_4$ and $J_5$ are fitted using the total energies
of four AFM structures. $\Delta$E are the energy differences
between the fitted energies from the Heisenberg model and 
the calculated ones. Unit of energy is in meV,
and the unit of exchange integrals are in meV/$\mu_B^2$.
 }
\begin{tabular}{ccrrr}
\hline \hline
SC & status& & Energy  & $\Delta$ E \\
\hline
$a$ & FM  & $E_0+16E_3+16E_4+4E_5$ &       0  & -179.84 \\
$b$ & FM  & $E_0-16E_3-16E_4+4E_5$ & -1355.66 & -158.22 \\
$c$ & FM  & $E_0-4E_5$             & -1004.93 & -8.41   \\
$d$ & FM  & $E_0$                  &  -910.32 & -19.78  \\
$e$ & AFM & $E_0+8E_3+8E_4$        &  -596.59 & 0       \\
$f$ & AFM & $E_0-8E_3-8E_4$    & -1263.61 & 0       \\
$g$ & AFM & $E_0-16E_4-4E_5$       & -1624.95 & 0       \\
$h$ & AFM & $E_0+16E_4-4E_5$       &  -407.72 & 0       \\
\hline
& & $J_3$=-0.45 \quad $J_4=-4.92$ & $J_5=-1.85$\\
\hline
\end{tabular}
\label{tab:diff_Js}
\end{center}
\end{table}

\appendix
\section{Ground state spin configurations}

\label{sec:appendix}

In this appendix, we discuss in more details about the magnetic structures of
TbMn$_2$O$_5$ and the (super)exchange interactions $J$s. \cite{chapon04}
To determine the $J$s, we calculate the total energies
of eight SCs shown in Fig. \ref{fig:spinconfigurations}, under the fixed
{\it high} symmetry structure $H$.
The total energies of these SCs are listed in Table \ref{tab:diff_Js}, 
among which SC $g$ has the lowest total energy. 
We then use a Heisenberg model with NN interaction,
\begin{equation}
E=E_0-\sum_{ij}J_{ij}\textbf{S}_i\cdot\textbf{S}_{j} \, ,
\label{eq:heisenbergmodel}
\end{equation}
to fit $J$s and to check whether the $g$ is indeed the the ground 
state SC. 
Here, ${\bf S}_i$ is the spin of the $i$-th Mn ion 
and $J_{ij}$ denotes the exchange integral between two NN
$i$-th and $j$-th atoms.
$J_3$, $J_4$ and $J_5$ are fitted using four AFM SCs, and 
the results are listed in Table \ref{tab:diff_Js}. 
Since $J_1$ and $J_2$ is not relevant here, they are not fitted.
We calculate the total energies of the four
FM SCs from Eq.(\ref{eq:heisenbergmodel}). These results are in 
reasonable good agreement
with the first-principles calculations, 
suggesting that the Heisenberg 
model is valid even in the FM phase when applied to TbMn$_2$O$_5$.
From Table \ref{tab:diff_Js}, we find all the three exchange integrals are
negative, i.e., of AFM type,
and $| J_3 | \ll | J_4 |, | J_5 |$.
Therefore the ground state SC are mainly determined by
$J_4$ and $J_5$. 
It is easy to see that SC $g$ (and $g'$)
indeed has the lowest energy in this model.

In both SCs $g$ and $g'$, Mn$^{4+}$ form an 
AFM square lattice in the $ab$ plane, whereas Mn$^{3+}$ couples to
Mn$^{4+}$ either antiferromagnetically via $J_4$ along the $a$ axis or
with alternating sign via $J_3$ along the $b$ axis.
The zigzag chains along the
$b$ axis can be simplified as
Mn$^{3+}$-Mn$^{4+}$-Mn$^{3+}$ segments linked via $J_5$ superexchange
interactions. Inside the segments Mn$^{3+}$ and Mn$^{4+}$ interact through
superexchange $J_3$, whereas the chains couple to each other
through $J_4$. 
The AFM chains along the $a$ axis are of two types, labeled as I and
II respectively in Fig. \ref{fig:chains} .
Adjacent chains I and II along the $b$ axis
couple to each other via $J_3$, 
in which half of them are frustrated.  

The SCs $g$ and $g'$ have several energetically
degenerate SCs as shown in Fig. \ref{fig:chains}.
There are two relative positions between
chain I and II, $l$ and $r$. ``$l$'' (``$r$'') means that the
Mn$^{4+}$ ions in chain I have the same spins to the  Mn$^{4+}$ ions on their 
{\it left} ({\it right})  side in chain II in each unit cell. 
Figure \ref{fig:chains} shows 4 different combinations
of chains I and II. Figure \ref{fig:chains}(a) and
(b) are the ground SC $g$ and $g'$ respectively, whereas
Fig.\ref{fig:chains}(c) and (d) are in fact 
two domain wall structures $dw$ and $dw'$.
The two domain walls are energetically degenerate to $g$ and $g'$ 
in the NN Heisenberg model, because the $J_3$ exchange interactions between
chain I and II cancel out no matter what their relative positions are.
However, after turning on the spin-lattice coupling, the crystal structure $H$
will relax to $L$ or $R$, and the energy degeneracy between the domain wall
structures ($dw$ and $dw'$) to SCs $g$ and $g'$
will be lifted.



\begin{thebibliography}{28}
\expandafter\ifx\csname natexlab\endcsname\relax\def\natexlab#1{#1}\fi
\expandafter\ifx\csname bibnamefont\endcsname\relax
  \def\bibnamefont#1{#1}\fi
\expandafter\ifx\csname bibfnamefont\endcsname\relax
  \def\bibfnamefont#1{#1}\fi
\expandafter\ifx\csname citenamefont\endcsname\relax
  \def\citenamefont#1{#1}\fi
\expandafter\ifx\csname url\endcsname\relax
  \def\url#1{\texttt{#1}}\fi
\expandafter\ifx\csname urlprefix\endcsname\relax\def\urlprefix{URL }\fi
\providecommand{\bibinfo}[2]{#2}
\providecommand{\eprint}[2][]{\url{#2}}

\bibitem[{\citenamefont{Kimura et~al.}(2003)\citenamefont{Kimura, Goto,
  Shintani, Ishizaka, Arima, and Tokura1}}]{kimura03}
\bibinfo{author}{\bibfnamefont{T.}~\bibnamefont{Kimura}},
  \bibinfo{author}{\bibfnamefont{T.}~\bibnamefont{Goto}},
  \bibinfo{author}{\bibfnamefont{H.}~\bibnamefont{Shintani}},
  \bibinfo{author}{\bibfnamefont{K.}~\bibnamefont{Ishizaka}},
  \bibinfo{author}{\bibfnamefont{T.}~\bibnamefont{Arima}}, \bibnamefont{and}
  \bibinfo{author}{\bibfnamefont{Y.}~\bibnamefont{Tokura1}},
  \bibinfo{journal}{Nature(Landon)} \textbf{\bibinfo{volume}{426}},
  \bibinfo{pages}{55} (\bibinfo{year}{2003}).

\bibitem[{\citenamefont{Goto et~al.}(2004)\citenamefont{Goto, Kimura, Lawes,
  Ramirez, and Tokura}}]{goto04}
\bibinfo{author}{\bibfnamefont{T.}~\bibnamefont{Goto}},
  \bibinfo{author}{\bibfnamefont{T.}~\bibnamefont{Kimura}},
  \bibinfo{author}{\bibfnamefont{G.}~\bibnamefont{Lawes}},
  \bibinfo{author}{\bibfnamefont{A.~P.} \bibnamefont{Ramirez}},
  \bibnamefont{and} \bibinfo{author}{\bibfnamefont{Y.}~\bibnamefont{Tokura}},
  \bibinfo{journal}{Phys. Rev. Lett.} \textbf{\bibinfo{volume}{92}},
  \bibinfo{pages}{257201} (\bibinfo{year}{2004}).

\bibitem[{\citenamefont{Hur et~al.}(2004{\natexlab{a}})\citenamefont{Hur, Park,
  Sharma, Ahn, Guha, and Cheong}}]{hur04}
\bibinfo{author}{\bibfnamefont{N.}~\bibnamefont{Hur}},
  \bibinfo{author}{\bibfnamefont{S.}~\bibnamefont{Park}},
  \bibinfo{author}{\bibfnamefont{P.~A.} \bibnamefont{Sharma}},
  \bibinfo{author}{\bibfnamefont{J.~S.} \bibnamefont{Ahn}},
  \bibinfo{author}{\bibfnamefont{S.}~\bibnamefont{Guha}}, \bibnamefont{and}
  \bibinfo{author}{\bibfnamefont{S.-W.} \bibnamefont{Cheong}},
  \bibinfo{journal}{Nature (London)} \textbf{\bibinfo{volume}{429}},
  \bibinfo{pages}{392} (\bibinfo{year}{2004}{\natexlab{a}}).

\bibitem[{\citenamefont{Chapon et~al.}(2004)\citenamefont{Chapon, Blake,
  Gutmann, Park, Hur, Radaelli, and Cheong}}]{chapon04}
\bibinfo{author}{\bibfnamefont{L.~C.} \bibnamefont{Chapon}},
  \bibinfo{author}{\bibfnamefont{G.~R.} \bibnamefont{Blake}},
  \bibinfo{author}{\bibfnamefont{M.~J.} \bibnamefont{Gutmann}},
  \bibinfo{author}{\bibfnamefont{S.}~\bibnamefont{Park}},
  \bibinfo{author}{\bibfnamefont{N.}~\bibnamefont{Hur}},
  \bibinfo{author}{\bibfnamefont{P.~G.} \bibnamefont{Radaelli}},
  \bibnamefont{and} \bibinfo{author}{\bibfnamefont{S.~W.}
  \bibnamefont{Cheong}}, \bibinfo{journal}{Phys. \ Rev. \ Lett.}
  \textbf{\bibinfo{volume}{93}}, \bibinfo{pages}{177402}
  (\bibinfo{year}{2004}).

\bibitem[{\citenamefont{Blake et~al.}(2005)\citenamefont{Blake, Chapon,
  Radaelli, Park, Hur, Cheong, and Rodriguez-Carvajal}}]{blake05}
\bibinfo{author}{\bibfnamefont{G.~R.} \bibnamefont{Blake}},
  \bibinfo{author}{\bibfnamefont{L.~C.} \bibnamefont{Chapon}},
  \bibinfo{author}{\bibfnamefont{P.~G.} \bibnamefont{Radaelli}},
  \bibinfo{author}{\bibfnamefont{S.}~\bibnamefont{Park}},
  \bibinfo{author}{\bibfnamefont{N.}~\bibnamefont{Hur}},
  \bibinfo{author}{\bibfnamefont{S.-W.} \bibnamefont{Cheong}},
  \bibnamefont{and}
  \bibinfo{author}{\bibfnamefont{J.}~\bibnamefont{Rodriguez-Carvajal}},
  \bibinfo{journal}{Phys. Rev. B} \textbf{\bibinfo{volume}{71}},
  \bibinfo{pages}{214402} (\bibinfo{year}{2005}).

\bibitem[{\citenamefont{Hur et~al.}(2004{\natexlab{b}})\citenamefont{Hur, Park,
  Sharma, Guha, and Cheong}}]{hur04b}
\bibinfo{author}{\bibfnamefont{N.}~\bibnamefont{Hur}},
  \bibinfo{author}{\bibfnamefont{S.}~\bibnamefont{Park}},
  \bibinfo{author}{\bibfnamefont{P.~A.} \bibnamefont{Sharma}},
  \bibinfo{author}{\bibfnamefont{S.}~\bibnamefont{Guha}}, \bibnamefont{and}
  \bibinfo{author}{\bibfnamefont{S.-W.} \bibnamefont{Cheong}},
  \bibinfo{journal}{Phys. Rev. Lett.} \textbf{\bibinfo{volume}{93}},
  \bibinfo{pages}{107207} (\bibinfo{year}{2004}{\natexlab{b}}).

\bibitem[{\citenamefont{Cheong and Mostovoy}(2007)}]{cheong07}
\bibinfo{author}{\bibfnamefont{S.-W.} \bibnamefont{Cheong}} \bibnamefont{and}
  \bibinfo{author}{\bibfnamefont{M.}~\bibnamefont{Mostovoy}},
  \bibinfo{journal}{Nature Materials} \textbf{\bibinfo{volume}{6}},
  \bibinfo{pages}{13} (\bibinfo{year}{2007}).

\bibitem[{\citenamefont{Fiebig}(2005)}]{fiebig05}
\bibinfo{author}{\bibfnamefont{M.}~\bibnamefont{Fiebig}}, \bibinfo{journal}{J.
  Phys. D: Appl. Phys.} \textbf{\bibinfo{volume}{286}}, \bibinfo{pages}{R123}
  (\bibinfo{year}{2005}).

\bibitem[{\citenamefont{Kagomiya et~al.}(2003)\citenamefont{Kagomiya,
  Matsumoto, Kohn, Fukuda, Shoubu, Kimura, Noda, and Ikeda}}]{kagomiya03}
\bibinfo{author}{\bibfnamefont{I.}~\bibnamefont{Kagomiya}},
  \bibinfo{author}{\bibfnamefont{S.}~\bibnamefont{Matsumoto}},
  \bibinfo{author}{\bibfnamefont{K.}~\bibnamefont{Kohn}},
  \bibinfo{author}{\bibfnamefont{Y.}~\bibnamefont{Fukuda}},
  \bibinfo{author}{\bibfnamefont{T.}~\bibnamefont{Shoubu}},
  \bibinfo{author}{\bibfnamefont{H.}~\bibnamefont{Kimura}},
  \bibinfo{author}{\bibfnamefont{Y.}~\bibnamefont{Noda}}, \bibnamefont{and}
  \bibinfo{author}{\bibfnamefont{N.}~\bibnamefont{Ikeda}},
  \bibinfo{journal}{Ferroelectrics} \textbf{\bibinfo{volume}{286}},
  \bibinfo{pages}{889} (\bibinfo{year}{2003}).

\bibitem[{\citenamefont{ValdesAguilar et~al.}(2006)\citenamefont{ValdesAguilar,
  Sushkov, Park, Cheong, , and Drew}}]{aguilar06}
\bibinfo{author}{\bibfnamefont{R.}~\bibnamefont{ValdesAguilar}},
  \bibinfo{author}{\bibfnamefont{A.~B.} \bibnamefont{Sushkov}},
  \bibinfo{author}{\bibfnamefont{S.}~\bibnamefont{Park}},
  \bibinfo{author}{\bibfnamefont{S.~W.} \bibnamefont{Cheong}}, ,
  \bibnamefont{and} \bibinfo{author}{\bibfnamefont{H.~D.} \bibnamefont{Drew}},
  \bibinfo{journal}{Phys. Rev. B} \textbf{\bibinfo{volume}{74}},
  \bibinfo{pages}{184404} (\bibinfo{year}{2006}).

\bibitem[{\citenamefont{Katsura et~al.}(2005)\citenamefont{Katsura, Nagaosa,
  and Balatsky}}]{katsura05}
\bibinfo{author}{\bibfnamefont{H.}~\bibnamefont{Katsura}},
  \bibinfo{author}{\bibfnamefont{N.}~\bibnamefont{Nagaosa}}, \bibnamefont{and}
  \bibinfo{author}{\bibfnamefont{A.~V.} \bibnamefont{Balatsky}},
  \bibinfo{journal}{Phys. Rev. Lett.} \textbf{\bibinfo{volume}{95}},
  \bibinfo{pages}{057205} (\bibinfo{year}{2005}).

\bibitem[{\citenamefont{Sergienko and Dagotto}(2006)}]{sergienko06}
\bibinfo{author}{\bibfnamefont{I.~A.} \bibnamefont{Sergienko}}
  \bibnamefont{and} \bibinfo{author}{\bibfnamefont{E.}~\bibnamefont{Dagotto}},
  \bibinfo{journal}{Phys. Rev. B.} \textbf{\bibinfo{volume}{73}},
  \bibinfo{pages}{094434} (\bibinfo{year}{2006}).

\bibitem[{\citenamefont{Alonso et~al.}(1997)\citenamefont{Alonso, Casais,
  Martinez-Lope, Martinez, and Dernandez-Diaz}}]{alonso97}
\bibinfo{author}{\bibfnamefont{J.~A.} \bibnamefont{Alonso}},
  \bibinfo{author}{\bibfnamefont{M.~T.} \bibnamefont{Casais}},
  \bibinfo{author}{\bibfnamefont{M.~J.} \bibnamefont{Martinez-Lope}},
  \bibinfo{author}{\bibfnamefont{J.~L.} \bibnamefont{Martinez}},
  \bibnamefont{and} \bibinfo{author}{\bibfnamefont{M.~T.}
  \bibnamefont{Dernandez-Diaz}}, \bibinfo{journal}{J. \ Phys.: \ Condens.\
  Matter} \textbf{\bibinfo{volume}{9}}, \bibinfo{pages}{8515}
  (\bibinfo{year}{1997}).

\bibitem[{hu0()}]{hu07}
\bibinfo{note}{J. Hu, arXiv:0705.0955}.

\bibitem[{\citenamefont{Chapon et~al.}(2006)\citenamefont{Chapon, Radaelli,
  Blake, Park, and Cheong}}]{chapon06}
\bibinfo{author}{\bibfnamefont{L.~C.} \bibnamefont{Chapon}},
  \bibinfo{author}{\bibfnamefont{P.~G.} \bibnamefont{Radaelli}},
  \bibinfo{author}{\bibfnamefont{G.~R.} \bibnamefont{Blake}},
  \bibinfo{author}{\bibfnamefont{S.}~\bibnamefont{Park}}, \bibnamefont{and}
  \bibinfo{author}{\bibfnamefont{S.-W.} \bibnamefont{Cheong}},
  \bibinfo{journal}{Phys. Rev. Lett.} \textbf{\bibinfo{volume}{96}},
  \bibinfo{pages}{097601} (\bibinfo{year}{2006}).

\bibitem[{\citenamefont{Wang et~al.}(2007)\citenamefont{Wang, Guo, and
  He}}]{wang07}
\bibinfo{author}{\bibfnamefont{C.}~\bibnamefont{Wang}},
  \bibinfo{author}{\bibfnamefont{G.-C.} \bibnamefont{Guo}}, \bibnamefont{and}
  \bibinfo{author}{\bibfnamefont{L.}~\bibnamefont{He}}, \bibinfo{journal}{Phys.
  Rev. Lett.}  (\bibinfo{year}{2007}).

\bibitem[{\citenamefont{Perdew et~al.}(1996)\citenamefont{Perdew, Burke, and
  Ernzerhof}}]{perdew96}
\bibinfo{author}{\bibfnamefont{J.~P.} \bibnamefont{Perdew}},
  \bibinfo{author}{\bibfnamefont{K.}~\bibnamefont{Burke}}, \bibnamefont{and}
  \bibinfo{author}{\bibfnamefont{M.}~\bibnamefont{Ernzerhof}},
  \bibinfo{journal}{Phys. \ Rev. \ Lett.} \textbf{\bibinfo{volume}{77}},
  \bibinfo{pages}{3865} (\bibinfo{year}{1996}).

\bibitem[{\citenamefont{Kresse and Hafner}(1993)}]{kresse93}
\bibinfo{author}{\bibfnamefont{G.}~\bibnamefont{Kresse}} \bibnamefont{and}
  \bibinfo{author}{\bibfnamefont{J.}~\bibnamefont{Hafner}},
  \bibinfo{journal}{Phys. Rev. B} \textbf{\bibinfo{volume}{47}},
  \bibinfo{pages}{RC558} (\bibinfo{year}{1993}).

\bibitem[{\citenamefont{Kresse and Furthmuller}(1996)}]{kresse96}
\bibinfo{author}{\bibfnamefont{G.}~\bibnamefont{Kresse}} \bibnamefont{and}
  \bibinfo{author}{\bibfnamefont{J.}~\bibnamefont{Furthmuller}},
  \bibinfo{journal}{Phys. Rev. B} \textbf{\bibinfo{volume}{54}},
  \bibinfo{pages}{11169} (\bibinfo{year}{1996}).

\bibitem[{\citenamefont{Blochl}(1994)}]{blochl94}
\bibinfo{author}{\bibfnamefont{P.~E.} \bibnamefont{Blochl}},
  \bibinfo{journal}{Phys. Rev. B} \textbf{\bibinfo{volume}{50}},
  \bibinfo{pages}{17953} (\bibinfo{year}{1994}).

\bibitem[{\citenamefont{Lines and Glass}(2001)}]{lines_book}
\bibinfo{author}{\bibfnamefont{M.~E.} \bibnamefont{Lines}} \bibnamefont{and}
  \bibinfo{author}{\bibfnamefont{A.~M.} \bibnamefont{Glass}},
  \emph{\bibinfo{title}{Principles and Applications of Ferroelectrics and
  Related Materials}} (\bibinfo{publisher}{Oxford University Press},
  \bibinfo{year}{2001}).

\bibitem[{\citenamefont{He et~al.}(2002)\citenamefont{He, Neaton, Cohen,
  Vanderbilt, and Homes}}]{he02}
\bibinfo{author}{\bibfnamefont{L.}~\bibnamefont{He}},
  \bibinfo{author}{\bibfnamefont{J.~B.} \bibnamefont{Neaton}},
  \bibinfo{author}{\bibfnamefont{M.~H.} \bibnamefont{Cohen}},
  \bibinfo{author}{\bibfnamefont{D.}~\bibnamefont{Vanderbilt}},
  \bibnamefont{and} \bibinfo{author}{\bibfnamefont{C.~C.} \bibnamefont{Homes}},
  \bibinfo{journal}{Phys. Rev. B} \textbf{\bibinfo{volume}{65}},
  \bibinfo{pages}{214112} (\bibinfo{year}{2002}).

\bibitem[{\citenamefont{Stokes and Hatch}(1999)}]{smodes}
\bibinfo{author}{\bibfnamefont{H.~T.} \bibnamefont{Stokes}} \bibnamefont{and}
  \bibinfo{author}{\bibfnamefont{D.~M.} \bibnamefont{Hatch}},
  \bibinfo{journal}{SMODES, www.physics.byu.edu/~stokesh/isotropy.html.}
  (\bibinfo{year}{1999}).

\bibitem[{\citenamefont{Mihailova et~al.}(2005)\citenamefont{Mihailova,
  Gospodinov, Guttler, Yen, Litvinchuk, and Iliev}}]{mihailova05}
\bibinfo{author}{\bibfnamefont{B.}~\bibnamefont{Mihailova}},
  \bibinfo{author}{\bibfnamefont{M.~M.} \bibnamefont{Gospodinov}},
  \bibinfo{author}{\bibfnamefont{B.}~\bibnamefont{Guttler}},
  \bibinfo{author}{\bibfnamefont{F.}~\bibnamefont{Yen}},
  \bibinfo{author}{\bibfnamefont{A.~P.} \bibnamefont{Litvinchuk}},
  \bibnamefont{and} \bibinfo{author}{\bibfnamefont{M.~N.} \bibnamefont{Iliev}},
  \bibinfo{journal}{Phys. Rev. B} \textbf{\bibinfo{volume}{71}},
  \bibinfo{pages}{172301} (\bibinfo{year}{2005}).

\bibitem[{\citenamefont{King-Smith and Vanderbilt}(1993)}]{king-smith93}
\bibinfo{author}{\bibfnamefont{R.~D.} \bibnamefont{King-Smith}}
  \bibnamefont{and}
  \bibinfo{author}{\bibfnamefont{D.}~\bibnamefont{Vanderbilt}},
  \bibinfo{journal}{Phys. Rev. B} \textbf{\bibinfo{volume}{47}},
  \bibinfo{pages}{1651} (\bibinfo{year}{1993}).

\bibitem[{\citenamefont{Betouras et~al.}(2007)\citenamefont{Betouras,
  Giovannetti, and den Brink}}]{betouras07}
\bibinfo{author}{\bibfnamefont{J.~J.} \bibnamefont{Betouras}},
  \bibinfo{author}{\bibfnamefont{G.}~\bibnamefont{Giovannetti}},
  \bibnamefont{and} \bibinfo{author}{\bibfnamefont{J.} \bibnamefont{van den
  Brink}}, \bibinfo{journal}{Phys. Rev. Lett.}  (\bibinfo{year}{2007}).

\bibitem[{\citenamefont{Sushkov et~al.}(2005)\citenamefont{Sushkov,
  Tchernyshyov, Ratcliff, Cheong, and Drew}}]{sushkov05}
\bibinfo{author}{\bibfnamefont{A.~B.} \bibnamefont{Sushkov}},
  \bibinfo{author}{\bibfnamefont{O.}~\bibnamefont{Tchernyshyov}},
  \bibinfo{author}{\bibfnamefont{W.}~\bibnamefont{Ratcliff}},
  \bibinfo{author}{\bibfnamefont{S.~W.} \bibnamefont{Cheong}},
  \bibnamefont{and} \bibinfo{author}{\bibfnamefont{H.~D.} \bibnamefont{Drew}},
  \bibinfo{journal}{Phys. Rev. Lett.} \textbf{\bibinfo{volume}{94}},
  \bibinfo{pages}{137202} (\bibinfo{year}{2005}).

\bibitem[{\citenamefont{Fennie and Rabe}(2006)}]{fennie06}
\bibinfo{author}{\bibfnamefont{C.~J.} \bibnamefont{Fennie}} \bibnamefont{and}
  \bibinfo{author}{\bibfnamefont{K.~M.} \bibnamefont{Rabe}},
  \bibinfo{journal}{Phys. Rev. Lett.} \textbf{\bibinfo{volume}{96}},
  \bibinfo{pages}{205505} (\bibinfo{year}{2006}).

\end{thebibliography}

\end{document}